\documentstyle[epsf]{article}

\font\tenrm=cmr10
\font\tenit=cmti10
\font\elevenbf=cmbx10 scaled\magstep 1
\font\elevenrm=cmr10 scaled\magstep 1
 1

\newcommand{\agt}{\mathrel{\raisebox{-.6ex}{$\stackrel{\textstyle>}{\sim}$}}}

\renewenvironment{thebibliography}[1]
 { \elevenrm
   \begin{list}{\arabic{enumi}.}
    {\usecounter{enumi} \setlength{\parsep}{0pt}
     \setlength{\itemsep}{3pt} \settowidth{\labelwidth}{#1.}
     \sloppy
    }}{\end{list}}

\begin{document}
\font\fortssbx=cmssbx10 scaled \magstep2
\hbox to \hsize{
\hskip.5in \raise.1in\hbox{\fortssbx University of Wisconsin - Madison}
\hfill\vbox{\hbox{\bf MAD/PH/803}
            \hbox{\bf RAL-93-085}
            \hbox{October 1993}} }

\begin{center}
\vglue 0.6cm
{
 {\elevenbf        \vglue 10pt
               RGE RESULTS FOR SUPERSYMMETRIC GUTS\footnote{Presented
by MSB at the XXIII International Symposium on Multiparticle  Dynamics,
12-17 September 1993, Aspen, Colorado.
}
\\}
\vglue 1.0cm
{\tenrm V.~Barger$^a$, M.~S.~Berger$^a$, P.~Ohmann$^a$, and
R.~J.~N.~Phillips$^b$\\}
\baselineskip=13pt
{\tenit $^a$Physics Department, University of Wisconsin\\}
\baselineskip=12pt
{\tenit Madison, WI 53706, USA\\}
\vglue 6pt
{\tenit $^b$Rutherford Appleton Laboratroy\\}
\baselineskip=12pt
{\tenit Chilton, Didcot, Oxon OX11 0QX, UK\\}}

\vglue 0.8cm
{\tenrm ABSTRACT}

\end{center}

\vglue 0.3cm
{\rightskip=3pc
 \leftskip=3pc
 \tenrm\baselineskip=12pt
 \noindent
The scaling behavior of gauge couplings and fermion Yukawa
couplings in the minimal supersymmetric model is discussed.
The relevance of the top quark Yukawa coupling fixed point in
establishing the top quark mass is described. The evolution of mixing angles
is presented.}

\vglue 0.4in
\baselineskip=14pt
\elevenrm

\vglue 0.6cm
{\elevenbf\noindent 1. Gauge Coupling Evolution}
\vglue 0.2cm
The one- and two-loop renormalization group equations (RGEs)
can be written for general Yukawa matrices as
\begin{eqnarray}
{{dg_i}\over {dt}}&=&{g_i\over{16\pi^2 }}\left [b_ig_i^2+{1\over {16\pi^2 }}
\left (\sum _{j=1}^3b_{ij}g_i^2g_j^2
-\sum _{j=U,D,E}a_{ij}g_i^2
{\bf Tr}[{\bf Y_j^{}Y_j^{\dagger }}]\right )\right ]\;,
\end{eqnarray}
with ${\bf Y_j^{}}\equiv {\bf U}$, ${\bf D}$, ${\bf E}$ (the Yukawa coupling
matrices) and $t=\ln \mu/M_G^{}$.
The low-energy values of the gauge couplings $g_1^{}$ and $g_2^{}$
lead
to a prediction for $g_3^{}$\cite{susygut} via the hypothesis of a grand
unified theory (GUT).
For the supersymmetric model with two Higgs doublets (MSSM),
the coefficients are given by\cite{EJ,susyrge2,bbo}
\begin{eqnarray}
b_i&=&({33\over 5},1,-3)\;, \quad
b_{ij} = \left( \begin{array}{c@{\quad}c@{\quad}c}
{199/25} & {27/5} & {88/5} \\
{9/5} & 25 & 24 \\ {11/5} & 9 & 14
\end{array} \right) \;, \quad
a_{ij} = \left( \begin{array}{c@{\quad}c@{\quad}c}
{26/5} & {14/5} & {18/5} \\
6 & 6 & 2 \\ 4 & 4 & 0
\end{array} \right) \;.
\end{eqnarray}
The two-loop gauge coupling terms involving the $b_{ij}$ affect the prediction
for $\alpha _3(M_Z^{})$
by $\approx 10\%$.
The two-loop Yukawa coupling terms involving the $a_{ij}$
affect the prediction for
$\alpha _3(M_Z^{})$ by $\approx 1\%$.
GUT scale threshold corrections can also affect the prediction for
$\alpha _3(M_Z^{})$\cite{hs}.

\vglue 0.6cm
{\elevenbf\noindent 2. Yukawa Coupling Evolution}
\vglue 0.2cm

At one-loop the particle content of the MSSM gives\cite{susyrge1}
\begin{eqnarray}
{{d{\bf U}}\over {dt}}&=&{1\over {16\pi ^2}}
\Big [-\sum c_ig_i^2+3{\bf UU^{\dagger }}+{\bf DD^{\dagger }}
+{\bf Tr}[3{\bf UU^{\dagger }}]\Big ]{\bf U} \;, \label{yuku} \\
{{d{\bf D}}\over {dt}}&=&{1\over {16\pi ^2}}
\Big [-\sum c_i^{\prime}g_i^2+3{\bf DD^{\dagger }}+{\bf UU^{\dagger }}
+{\bf Tr}[3{\bf DD^{\dagger }}+{\bf EE^{\dagger }}]\Big ]{\bf D} \;,
\label{yukd} \\
{{d{\bf E}}\over {dt}}&=&{1\over {16\pi ^2}}
\Big [-\sum c_i^{\prime \prime}g_i^2+3{\bf EE^{\dagger }}
+{\bf Tr}[3{\bf DD^{\dagger }}+{\bf EE^{\dagger }}]\Big ]{\bf E} \;,
\label{yuke}
\end{eqnarray}
where $c_i^{}=(13/15,3,16/3)$, $c'_i=(7/15,3,16/3)$, $c''_i=(9/5,3,0)$.
The two-loop equations in their full matrix form can be found in the
appendix of Ref.~\cite{bbo}.
The individual terms in these equations can be understood independently.
The terms involving the gauge couplings arise from the contribution
$c_i^{}(f)$ to the anomalous dimension of each field in the Yukawa coupling.
For example, $c_i^{}=c({q_L^{}})+c({u_R^{}})+c({H_2^{}})$ where
$c_i^{}(f)$ is ${{N^2-1}\over {N}}$ (0) for the fundamental representation
(singlet) of $SU(N)$ and
${3\over {10}}Y^2$ (suitably normalized so that $Y_{\tau }=2$)
for $U(1)_Y^{}$. Furthermore the trace contribution must arise from
fermion loops as in Figure 1.
\begin{center}
\epsfxsize=2in
\hspace{0in}
\epsffile{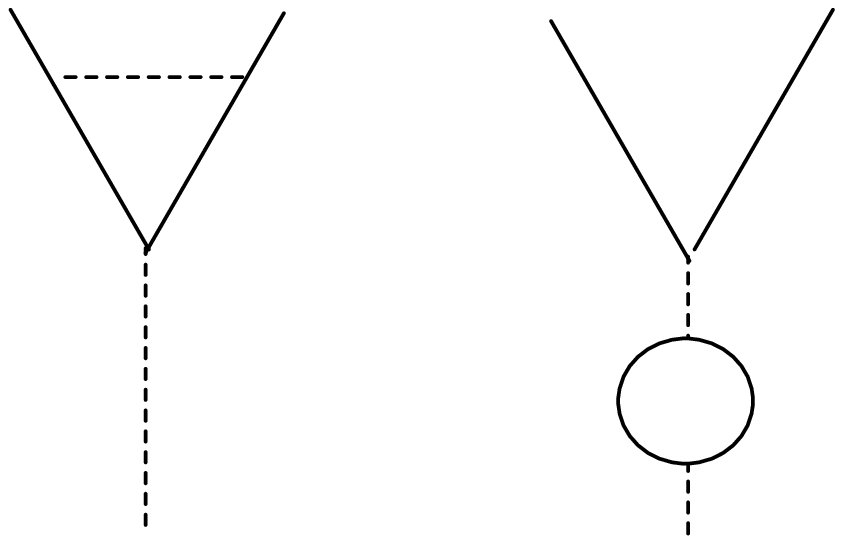}

\parbox{5.5in}{\small \small Fig.~1. Two of the
diagrams which contribute to the one-loop Yukawa
coupling renormalization group equations.}
\end{center}

We say that a variable $X$ scales when it obeys a differential equation of the
form
\begin{eqnarray}
{{dX}\over {dt}}&=&{X\over {16\pi ^2}}\Big [\dots \Big ]\;.
\end{eqnarray}
The gauge and Yukawa couplings are of this form to leading order in the fermion
hierarchy.
The scaling factors $S_i$ for the fermion evolution may be defined as
\begin{eqnarray}
\lambda _i(M_G^{})&=&S_i\lambda _i(m_t)\;,
\end{eqnarray}
The $S_i$  are plotted versus $\tan \beta $ for $m_t=150$ GeV in Figure 2.

\begin{center}
\epsfxsize=5.4in
\hspace{0in}
\epsffile{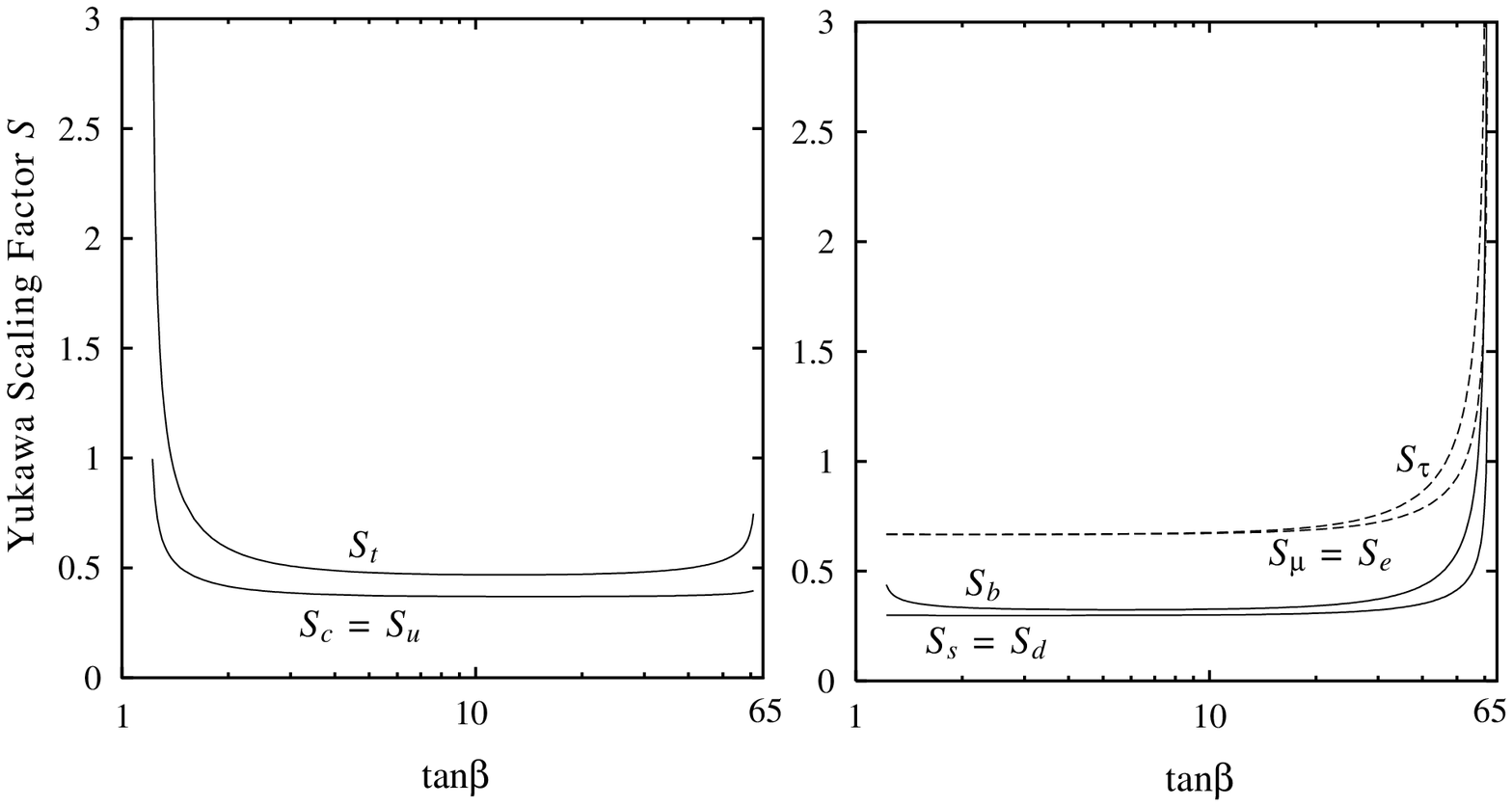}

\parbox{5.5in}{\small \center Fig.~2. Typical Yukawa scaling factors $S_i=
          \lambda_i(M_G)/\lambda_i(m_t)$ with $m_t=150$ GeV.}
\end{center}

The evolution of the gauge and Yukawa couplings (which are dimensionless
parameters) does not depend on the soft-supersymmetry breaking parameters
(which are dimensionful).
The analysis decouples to first order from the details of SUSY
breaking
though the
SUSY spectrum can still affect results through threshold effects.
The evolution of the soft-supersymmetry breaking parameters
do depend on dimensionless gauge and Yukawa couplings, however.
For instance the right-handed soft
top-squark mass $M_{\tilde{t}_R}$ has the RGE
\begin{eqnarray}
{{dM_{\tilde{t}_R}^2}\over {dt}}&=&{2 \over {16\pi ^2}}
\Big (-{16\over 15}g_1^2M_1^2-{16\over 3}g_3^2M_3^2+2\lambda _t^2X_t
\Big )\;,
\end{eqnarray}
where $X_t=M_{Q_L}^2+M_{t _R}^2+M_{H_2}^2+A_t^2$ is a combination of SUSY mass
parameters.
These soft-supersymmetry breaking parameters do not exhibit scaling
(the supersymmetric Higgs mass parameter $\mu $ appearing in the
superpotential does scale, however).

Grand unified theories give the boundary conditions to the above differential
equations. The grand unified group guarantees certain relations between
Yukawa couplings when the Higgs sector is required to be simple. The first
such example was $\lambda _b=\lambda _{\tau }$
given by Chanowitz, Ellis, and Gaillard\cite{ceg} in 1977.
Georgi and Jarlskog\cite{gj} subsequently
proposed viable relations for the lightest two
generations: $3\lambda _s=\lambda _{\mu},\;\;{1\over 3}\lambda _d=\lambda _e$.

\vglue 0.6cm
{\elevenbf\noindent 3. Fixed Points}
\vglue 0.2cm
Yukawa couplings if large are driven to a fixed point at the electroweak
scale. The Yukawa couplings are related to the fermions masses (in our
convention) by
\begin{eqnarray}\label{lambda_b}
\lambda_b(m_t) = {\sqrt2\, m_b(m_b)\over\eta_b v\cos\beta}\,, \qquad
\lambda_\tau(m_t) = {\sqrt2m_\tau(m_\tau)\over \eta_\tau v\cos\beta}\,, \qquad
\lambda_t(m_t) = {\sqrt2 m_t(m_t)\over v\sin\beta}\;.
\end{eqnarray}
The scaling factors $\eta _b$ and $\eta _{\tau }$ relate the Yukawa couplings
to their values at the scale $m_t$.
The evolution of these Yukawa couplings can be deduced from
Eqs.~(\ref{yuku}-\ref{yuke}),
\begin{eqnarray}
{{d\lambda _t}\over {dt}}&=&{{\lambda _t}\over {16\pi^2}}\left [
-{13\over 15}g_1^2-3g_2^2-{16\over 3}g_3^2+6\lambda _t^2+\lambda _b^2\right ]
\;, \\
{{dR_{b/\tau}}\over {dt}}&=&{{R_{b/\tau}}\over {16\pi^2}}\left [
-{4\over 3}g_1^2-{16\over 3}g_3^2+\lambda _t^2+3\lambda _b^2
-3\lambda _{\tau }^2\right ]\;.
\end{eqnarray}
where we have defined
$R_{b/\tau}\equiv {{\lambda _b}\over {\lambda _{\tau}}}$.
The behavior of the top quark Yukawa coupling is exhibited in Figure 3
assuming that the bottom quark and the tau lepton Yukawa couplings are
approximately unified at a SUSY-GUT scale of approximately $2\times 10^{16}$
GeV.
The coupling approaches a quasi-infrared fixed point\cite{pendleton}
of approximately $1.1$.
We have taken different GUT scale threshold corrections for each curve.
\begin{center}
\epsfxsize=4.75in
\hspace*{0in}
\epsffile{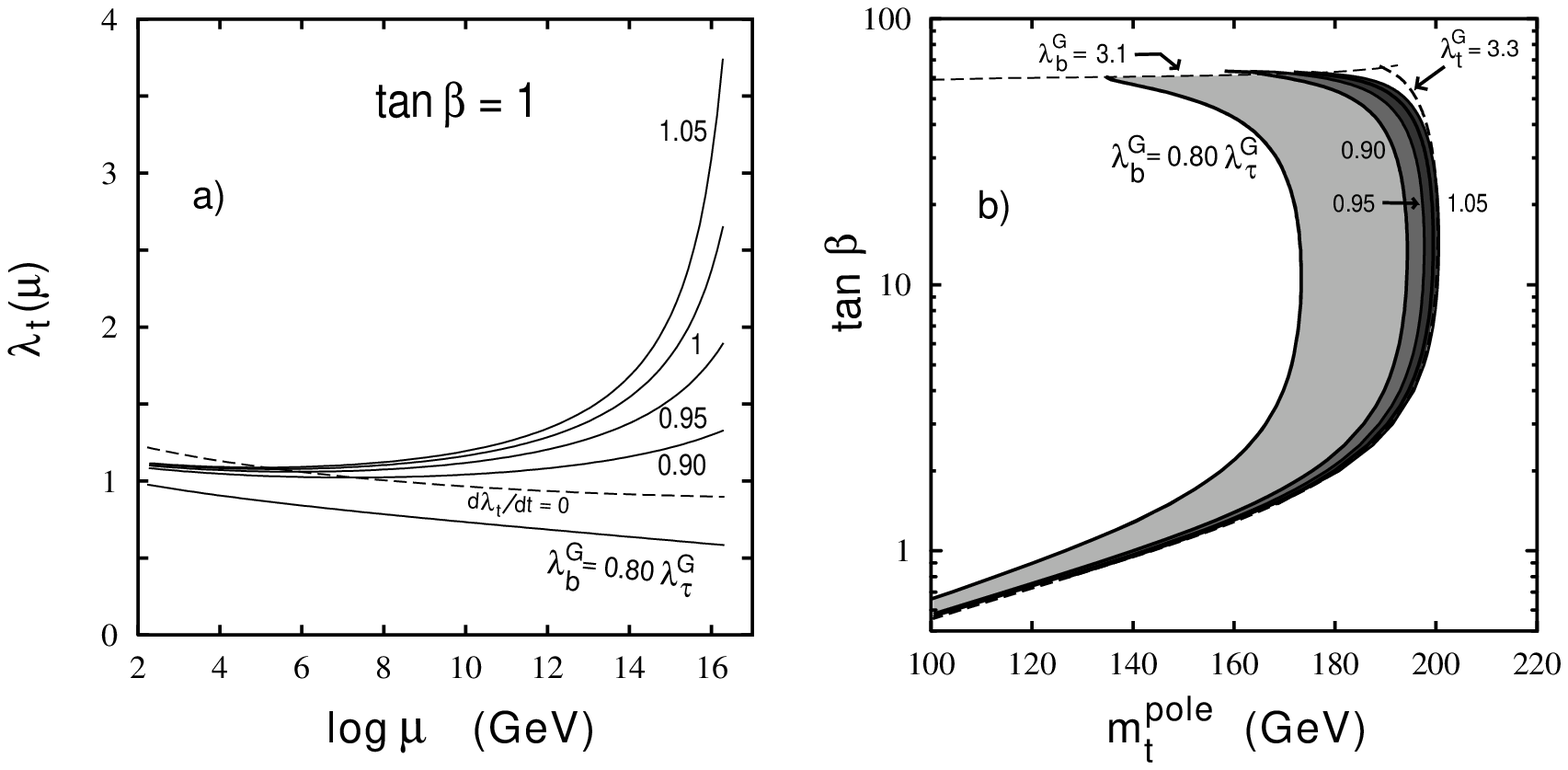}

\parbox{5.5in}{\small Fig.~3. a) The Yukawa coupling $\lambda _t$ approaches
a fixed point at the electroweak scale. All curves are attracted to the dashed
line as the scale $\mu $ is decreased. b) Effects of
GUT threshold corrections to Yukawa
coupling unification. Here $\alpha _3(M_Z^{})=0.118$ is assumed.}
\end{center}

For
$\lambda _b^G=0.80\lambda _{\tau}^G$, the fixed point does not adequately
describe the electroweak scale value of $\lambda _t$. This can be seen in
Figure 3b by looking at the $\tan \beta =1$ solutions in the $m_t,\tan \beta $
plane. For $\lambda _b^G=0.80\lambda _{\tau}^G$, the predicted value of $m_t$
is 20 GeV lower than the fixed point value. For larger values of
$\alpha _3(M_Z^{})$ still larger threshold corrections would be necessary to
avoid the fixed point.

The fixed point solution\cite{bbo,fkm,bbhz,cpw,bbop,lp} leads to the following
relation between the $\overline{DR}$ (dimensional reduction with minimal
subtraction) top quark mass and $\tan \beta $
\begin{eqnarray}
\lambda_t(m_t) &=& {\sqrt2 m_t(m_t)\over v\sin\beta} \qquad \Rightarrow
\qquad m_t(m_t)\approx {v\over {\sqrt{2}}}\sin \beta=(192 {\rm GeV})
\sin \beta \;,
\end{eqnarray}
Converting this relation for the top quark pole mass yields
\begin{eqnarray}
m_t^{\rm pole }&=&(200 {\rm GeV})\sin \beta \;.
\end{eqnarray}

If one takes the $\lambda _t$
fixed point solution seriously and also assumes that the
top quark mass $m_t^{\rm pole}$ is less than about 160 GeV, important
consequences result for
the Higgs sector of the MSSM. From Figure 3b it is clear that given these
assumptions $\tan \beta $ is very near one. Since $\tan \beta = 1$
is a flat direction
in the Higgs potential, the tree level mass is very small and the true mass
of the lightest
Higgs is given almost entirely by the one-loop radiative corrections.
This case was discussed in detail by Diaz and Haber\cite{dh}. The upper bound
that results is shown by the boundary of
the theoretically disallowed region in Figure 4. We have made
the assumption that colored SUSY particles are about 1 TeV to calculate the
radiative corrections.

\begin{center}
\epsfxsize=3.75in
\hspace{0in}
\epsffile{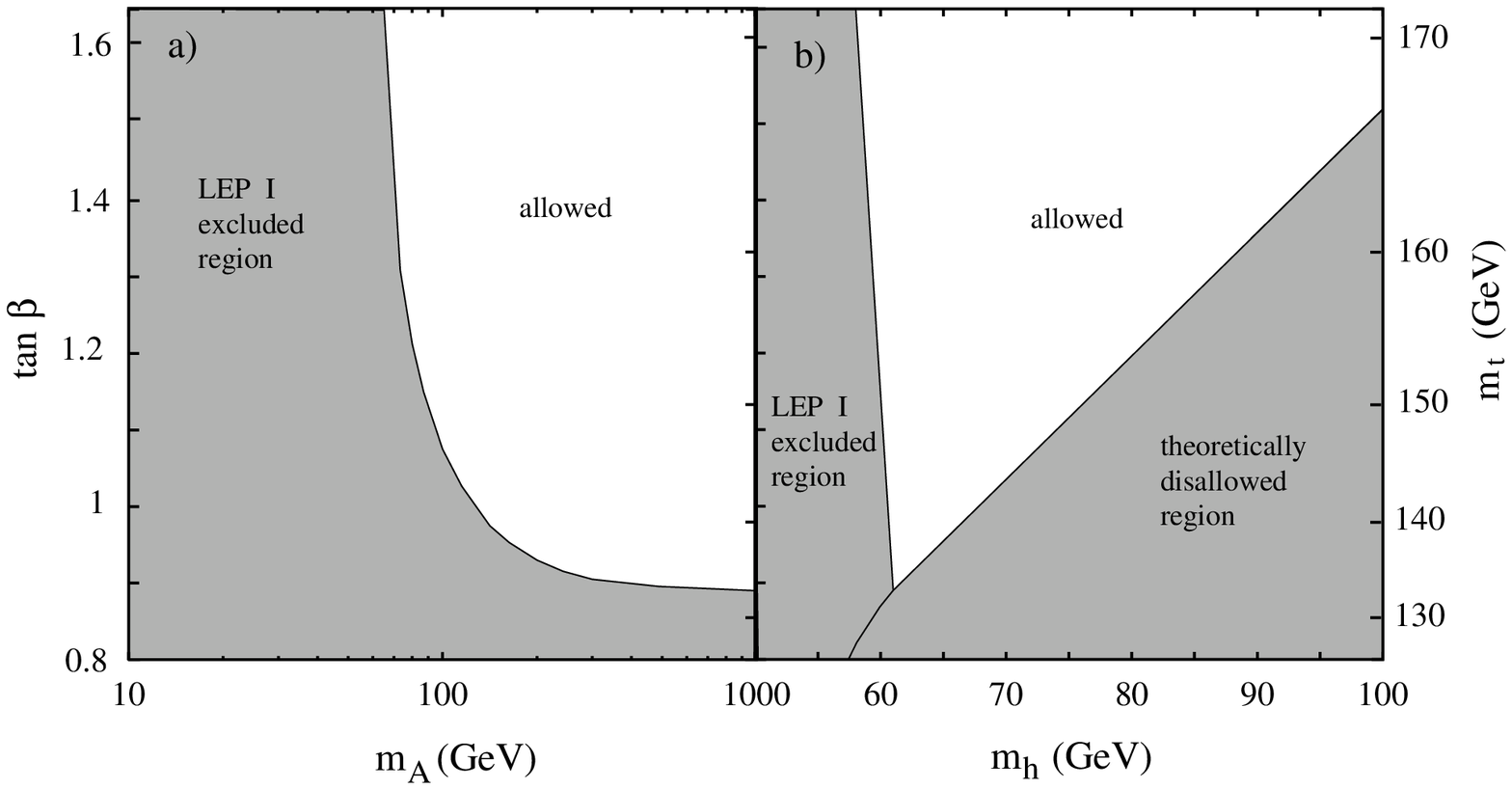}

\parbox{5.5in}{\small Fig.~4. The $\lambda_t$-fixed-point
solution regions allowed
by the LEP\,I data: (a)~in the $(m_A, \tan \beta )$ plane, (b)~in the $(m_h,
\tan \beta )$ plane. The top quark masses are $m_t({\rm pole})$, correlated to
$\tan \beta $ by the fixed point solution\cite{bbop}.}
\end{center}

LEP II will be able to discover the lightest SUSY Higgs boson
for $m_t^{\rm pole}$ up to 160 GeV
provided the fixed point solution for the top
Yukawa coupling is satisfied (see Figure 5).
If $m_t\agt 170$ GeV, $\tan \beta $ is not constrained.

\begin{center}
\epsfxsize=2.6in
\hspace{0in}
\epsffile{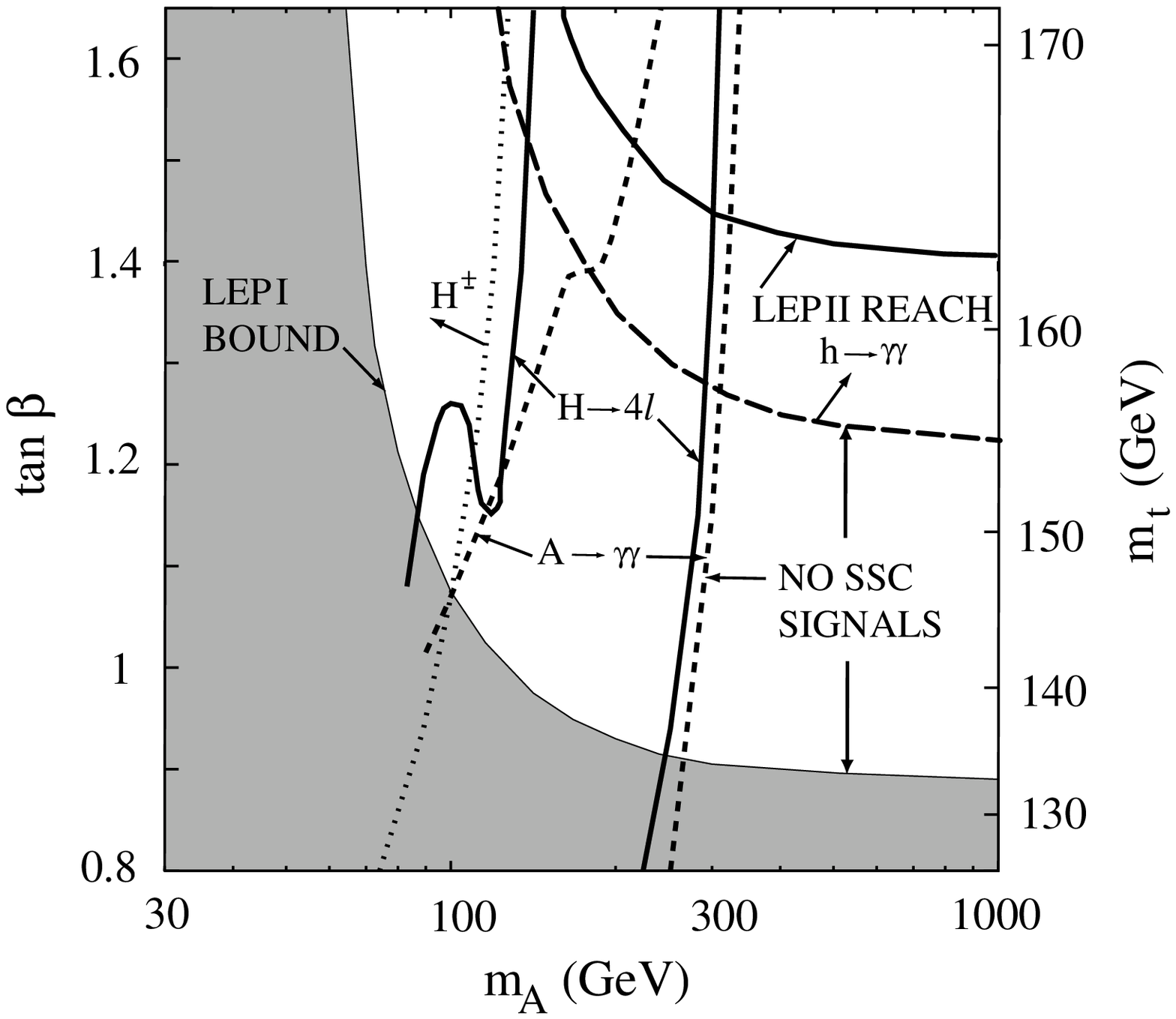}

\parbox{5.5in}{\small Fig.~5. Signal detectability regions, compared
with the LEP\,I allowed region of $\lambda_t$-fixed-point solutions
and the  probable reach of LEP\,II\cite{bbop}.
   The top quark masses are $m_t({\rm pole})$.}
\end{center}

\vglue 0.6cm
{\elevenbf\noindent 4. Universal Scaling of the CKM Matrix}
\vglue 0.2cm

There are three ``kinds'' of CKM matrix elements in regards to scaling
behavior:
(1) diagonal,
(2) mixing - between heavy and light generations, e.g. $V_{cb}$,
(3) mixing - between two light generations, e.g. $V_{us}$.
The evolution equations for the Yukawa couplings lead immediately to
evolution equations for the mixing angles in the CKM matrix:
\begin{eqnarray}
{{d{\bf U}}\over {dt}}\;,{{d{\bf D}}\over {dt}}
\Rightarrow {{dV_{\rm CKM}^{}}\over {dt}}\;,
\end{eqnarray}

Provided the mixings between heavy and light generations
are small one can prove that there are
only two types of scaling to leading order in the fermion
hierarchy\cite{op,bbo2,bs}
\begin{eqnarray}
{{d|V_1|^2}\over {dt}}&=&{{d|V_3|^2}\over {dt}}=0\;, \\
{{d|V_2|^2}\over {dt}}&=&-{{|V_2|^2}\over {8\pi^2}}\left (a_d\lambda _t^2
+a_u\lambda _b^2\right )\;\;{\rm +\:\:\:2-loop}\;.
\end{eqnarray}
We highlight some features of the evolution:
\begin{itemize}
\item Gauge couplings contributions
do not appear in the RGEs of the CKM elements.

\item The approximation of scaling is particularly good even though
 the Cabibbo angle
$|V_{us}|$ is not small.

\item The scaling behavior is a property of the hierarchy; it can be proven
to all orders in perturbation theory.

\item The universality of the scaling is model-independent. However
the amount of
the scaling varies between various models,
For example in the MSSM one has $a_u=a_d=1$ while in the Standard Model
$a_u=a_d=-{3\over 2}$.

\end{itemize}

The evolution of the CKM matrix is important when one examines the relations
between masses and mixings. One example that has been thoroughly examined
recently\cite{bbo,hrr,fls,dhr,adhrs} is
\begin{eqnarray}
|V_{cb}|&=&X\sqrt{{\lambda _c}\over {\lambda _t}}\quad {\rm at}\; M_G^{}\;.
\end{eqnarray}
where $X$ can account for GUT scale threshold corrections, or for Clebsch
factors\cite{adhrs}. Some other relations also occur under rather general
assumptions\cite{hr}.

\vglue 0.6cm
{\elevenbf\noindent 5. Acknowledgements}
\vglue 0.2cm
This research was supported
in part by the University of Wisconsin Research Committee with funds granted by
the Wisconsin Alumni Research Foundation, in part by the U.S.~Department of
Energy under contract no.~DE-AC02-76ER00881, and in part by the Texas National
Laboratory Research Commission under grant nos.~RGFY93-221 and FCFY9302.
MSB was supported in
part by an SSC Fellowship. PO was supported in
part by an NSF Graduate Fellowship.

\vglue 0.6cm
{\elevenbf\noindent 6. References}
\vglue 0.2cm


\begin{thebibliography}{00}
\frenchspacing

\bibitem{susygut} U.~Amaldi, W.~de~Boer, and H.~Furstenau,
Phys.\ Lett.\ {\bf B260} (1991) 447; J.~Ellis, S.~Kelley and
D.~V.~Nanopoulos, Phys.\ Lett.\ {\bf B260} (1991) 131;
P.~Langacker and M.~Luo, Phys.\ Rev.\ {\bf D44} (1991) 817 (1991).

\bibitem{EJ} M.~B.~Einhorn and D.~R.~T.~Jones, Nucl.\ Phys.\ {\bf196},
475 (1982).

\bibitem{susyrge2} J.~E.~Bj\"{o}rkman and D.~R.~T.~Jones, Nucl.\ Phys.\
{\bf B259} (1985) 533.

\bibitem{bbo}V.~Barger, M.S.~Berger, and P.~Ohmann, Phys.\ Rev.\ {\bf D47}
(1993) 1093.

\bibitem{hs} R.~Barbieri and L.~J.~Hall, Phys.\ Rev.\ Lett.\ {\bf 68},
752 (1992); P.~Langacker and N.~Polonsky, Phys.\ Rev.\ {\bf D47} (1992) 4028;
A.~E.~Faraggi, B.~Grinstein, S.~Meshkov, Phys.\ Rev.\ {\bf D47},
(1993) 5018;
L.~J.~Hall and U.~Sarid, Phys.\ Rev.\ Lett.\ {\bf 70} (1993) 2673;
K.~Hagiwara and Y.~Yamada, Phys.\ Rev.\ Lett.\ {\bf 70},
709 (1993).

\bibitem{susyrge1} K.~Inoue, A.~Kakuto, H.~Komatsu and S.~Takeshita,
Prog.\ Theor.\ Phys.\ {\bf 67} (1982) 1889.

\bibitem{ceg}M.~Chanowitz, J.~Ellis, and M.~K.~Gaillard, Nucl.\ Phys.\
{\bf B128} (1977) 506.

\bibitem{gj}H.~Georgi and C.~Jarlskog, Phys. Lett. {\bf 86B} (1979) 297.

\bibitem{pendleton} B.~Pendleton and G.~G.~Ross, Phys.\ Lett.\ {\bf 98B}
(1981) 291; C.~T.~Hill, Phys.\ Rev.\ {\bf D24} (1981) 691.

\bibitem{fkm} C.~D.~Froggatt, I.~G.~Knowles, and R.~G.~Moorhouse, Phys.\
Lett.\ {\bf B249}, 273 (1990); {\bf B298} (1993) 356.

\bibitem{bbhz} V.~Barger, M.~S.~Berger, T.~Han, and M.~Zralek, Phys.\ Rev.\
Lett.\ {\bf68} (1992) 3394.

\bibitem{cpw} M.~Carena, S.~Pokorski, and C.~E.~M.~Wagner, Nucl.\ Phys.\
{\bf B406} (1993) 59; W.~Bardeen, M.~Carena, S.~Pokorski, and C.~E.~M.~Wagner,
Munich preprint MPI-Ph/93-58.

\bibitem{bbop}V.~Barger et al., Phys.\ Lett.\ {\bf B314} (1993) 351.

\bibitem{lp} P.~Langacker and N.~Polonsky, University of Pennsylvania preprint
UPR-0556-T (1993).

\bibitem{dh} M.~Diaz and H.~Haber, Phys.\ Rev.\ {\bf D46} (1992) 3086.

\bibitem{op} M.~Olechowski and S.~Pokorski, Phys. Lett. {\bf B257} (1991) 388.

\bibitem{bbo2}V.~Barger, M.S.~Berger and P.~Ohmann, Phys. Rev. {\bf D47} (1993)
2038.

\bibitem{bs} K.~S.~Babu and Q.~Shafi, Phys.\ Rev.\ {\bf D47} (1993) 5004.

\bibitem{hrr} J.~Harvey, P.~Ramond, and D.~B.~Reiss, Phys.\ Lett.\ {\bf92B},
(1980) 309; Nucl.\ Phys.\ {\bf B199} (1982) 223.

\bibitem{fls} E.~M.~Freire, G.~Lazarides, and Q.~Shafi, Mod.\ Phys.\
{\bf A5} (1990) 2453.

\bibitem{dhr} S.~Dimopoulos, L.~Hall and S.~Raby, Phys.\ Rev.\ Lett. {\bf 68},
(1992) 1984; Phys.\ Rev.\ {\bf D45} (1992) 4192; {\bf D46} (1992) 4793;
G.W.~Anderson et al., Phys. Rev. {\bf D47} (1993) 3702.

\bibitem{adhrs} G.~Anderson, S.~Dimopoulos, L.~J.~Hall, S.~Raby, and
G.~D.~Starkman, Lawrence Berkeley Laboratory preprint LBL-33531 (1993).

\bibitem{hr} L.~J.~Hall and A.~Rasin, Phys.\ Rev.\ {\bf B315} (1993) 164.

\end{thebibliography}
\end{document}